\documentstyle[preprint,aps]{revtex}
\def\be{\begin{equation}}
\def\ee{\end{equation}}
\def\bea{\begin{eqnarray}}
\def\eea{\end{eqnarray}}

\begin{document}
\draft
\preprint{IP/BBSR/95-9 }
\title {\bf HIDDEN SYMMETRIES OF TWO DIMENSIONAL STRING
EFFECTIVE ACTION }
\author{ Jnanadeva Maharana \footnote{e-mail: maharana@iopb.ernet.in}}
\address{Institute of Physics, Bhubaneswar-751 005,
INDIA.}
\maketitle
\begin{center}
January 30, 1995
\end{center}
\begin{abstract}

The ten dimensional heterotic string effective action with
graviton, dilaton and antisymmetric tensor fields is
dimensionally reduced to two spacetime dimensions. The
resulting theory, with some constraints on backgrounds, admits
infinite sequence of conserved nonlocal currents. It is shown
that generators of the infinitesimal transformations associated
with these currents satisfy Kac-Moody algebra.

\end{abstract}
\vspace{2 cm}
 January 30, 1995
\narrowtext

\newpage


\par The purpose of this investigation is to unravel hidden
symmetries of dimensionally reduced string effective action in
two spacetime dimensions. Recently, we have shown [1] the existence
of an infinite set of nonlocal conserved currents (NCC) for the
reduced action with some constraints. The starting point is to
consider the heterotic string effective action in ten dimensions
with massless backgrounds such as graviton, dilaton and
antisymmetric tensor fields. Then, one toroidally compactifies
$d$ of its {\it internal} coordinates and requires that the
backgrounds are independent of these $d$ coordinates. It has been
demonstrated that the dimensionally reduced effective action is
invariant under global noncompact $O(d,d)$ symmetry
transformations [2,3].  Thus in $1+1$
dimensions the group is $O(8,8)$, and its algebra is denoted by
$\cal G$. The infinite sequence of currents
were  derived for this
action with some restrictions on the backgrounds. It is well
known that Kac-Moody algebra is intimately connected with integrable
systems, theories that admit NCC and string theory [4].

\par We exhibit  the
infinite parameter Lie algebra responsible for the NCC to be the
affine Kac-Moody algebra. First, it is shown, following the work
of Dolan and Roos [5], that there is an infinitesimal symmetry
transformation, associated with each of these currents, which
leave the Lagrangian invariant up to a total derivative term [6].
Then, the existence of the Kac-Moody algebra is proved, for the
problem at hand, by suitably adopting the remarkable result of
Dolan [7], derived for loop space and two dimensional chiral models.
We identify the
infinite parameter Lie algebra, crucial for the NCC, to be the
affine Kac-Moody subalgebra $C[\xi] \otimes {\cal G}$ following
ref.7. Here $C[\xi] \otimes {\cal G}$ is an infinite dimensional
Lie algebra defined over a ring of polynomials in the complex
variable $\xi$. A simple representation of the generators of
the algebra $C[\xi] \otimes {\cal G}$ is, ${\cal M}^{(n)}_{\alpha} =
T_{\alpha} {\xi}^{n}$, where $\{T_{\alpha} \} $ are the
generators of the finite parameter algebra ${\cal G}$, and $ n =
1, 2, ...\infty$. The generators of $C[\xi] \otimes {\cal G}$
satisfy $ [{\cal M}^{(n)}_{\alpha},{\cal M}^{(m)}_{\beta} ] = f_{\alpha
\beta \gamma}{\cal M}^{(m+n)}_{\gamma}$, when the algebra of the
generators of ${\cal G}$ is $[ T_{\alpha}, T_{\beta} ] = f_{\alpha \beta
\gamma} T_{\gamma}$ and $f_{\alpha \beta \gamma}$ are the
structure constants antisymmetric in their  indices and satisfy
the Jacobi identity.

\par  In what follows, we recapitulate the results of ref.2.
The  effective action in
 $\hat D=D+d$  dimensions ($\hat D=10$ for the present  case ) is,

\bea \hat S = \int d^{\hat D}x~ \sqrt{- \hat g} e^{-\hat\phi}
\big [\hat R
(\hat g) + \hat g^{\hat \mu \hat \nu} \partial_{\hat \mu} \hat\phi
\partial_{\hat \nu} \hat\phi  - {1 \over 12} ~ \hat H_{\hat \mu \hat \nu
\hat \rho} ~
\hat H^{\hat \mu \hat \nu \hat \rho}\big ].\eea
Note that $\hat S$ is the bosonic part of the heterotic string
effective action in critical dimension.
\noindent $\hat H$ is the field strength of antisymmetric tensor and $\hat
\phi$ is the dilaton. Here  all the  field
backgrounds to have been set to zero. We consider the theory in
a spacetime $M
\times K$, where $M$ is $D$ dimensional spacetime and the coordinates on
$M$ are denoted by $x^{\mu}$. The internal space, $K$, is $d$ dimensional and
$\{y^{\alpha}\}$, $\alpha = 1,2,..d$, are the coordinates. When the
backgrounds  are independent of
 $y^{\alpha}$ and the internal space is
taken to be torus, the metric $\hat g_{\hat \mu \hat \nu} $
 can be decomposed as

\bea \hat g_{\hat \mu \hat \nu} = \left (\matrix {g_{\mu \nu} +
A^{(1)\gamma}_{\mu} A^{(1)}_{\nu \gamma} &  A^{(1)}_{\mu \beta}\cr
A^{(1)}_{\nu \alpha} & G_{\alpha \beta}\cr}\right ),\eea

\noindent where $G_{\alpha \beta}$ is the internal metric and $g_{\mu\nu}$,
the $D$-dimensional space-time metric, depend on the coordinates $x^{\mu}$.
The dimensionally reduced action is,

\bea  S_{D}&= &\int d^Dx \sqrt {-g} e^{-\phi}
\bigg\{ R + g^{\mu \nu}
\partial_{\mu} \phi \partial_{\nu} \phi -{1\over 12}H_{\mu \nu \rho}
H^{\mu \nu \rho}\nonumber\\
& +& {1 \over 8} {\rm tr} (\partial_\mu M^{-1} \partial^\mu
M)- {1 \over 4}
{\cal F}^i_{\mu \nu} (M^{-1})_{ij} {\cal F}^{\mu \nu j} \bigg\}.
 \eea

\noindent Here $\phi=\hat\phi-{1\over 2}\log\det G$ is the
shifted dilaton.

\bea H_{\mu \nu \rho} = \partial_\mu B_{\nu \rho} - {1 \over 2} {\cal A}^i_\mu
\eta_{ij} {\cal F}^j_{\nu \rho} + ({\rm cyc.~ perms.}),\eea

\noindent ${\cal F}^i_{\mu \nu}$ is the $2d$-component vector of field
strengths
\bea {\cal F}^i_{\mu \nu} = \pmatrix {F^{(1) \alpha}_{\mu \nu}
\cr F^{(2)}_{\mu \nu \alpha}\cr} = \partial_\mu {\cal A}^i_\nu - \partial_\nu
{\cal A}^i_\mu \,\, ,\eea
\noindent $A^{(2)}_{\mu \alpha} = \hat B_{\mu \alpha} + B_{\alpha \beta}
A^{(1) \beta}_{\mu}$ (recall $B_{\alpha \beta}=\hat B_{\alpha \beta}$), and
the $2d\times 2d$ matrices $M$ and $\eta$ are defined as
\bea M = \pmatrix {G^{-1} & -G^{-1} B\cr
BG^{-1} & G - BG^{-1} B\cr},\qquad \eta =  \pmatrix {0 & 1\cr 1 & 0\cr}
\, .\eea

\noindent The action (3) is invariant under a global $O(d,d)$ transformation,

\bea M \rightarrow \Omega^T M \Omega, \qquad \Omega \eta \Omega^T = \eta,
\qquad
{\cal A}_{\mu}^i \rightarrow \Omega^i{}_j {\cal A}^j_\mu, \qquad {\rm
where} \qquad
\Omega \in O(d,d). \eea
\noindent and the shifted dilaton, $\phi$, remains invariant
under the $O(d,d)$ transformations.
Note that $M\in O(d,d)$ also and $M^T\eta M=\eta$. Thus if we
solve for a set of backgrounds, $M$,$\cal F$ and $\phi$,
satisfying the equations of motion they correspond to a vacuum
configuration of the string theory.

\par Let us consider the reduced action, eq.(3), in $1+1$
dimension. Note that
$H_{\mu \nu \rho} H^{\mu \nu \rho} $ term does not contribute to the
action in two spacetime dimensions. Moreover, we assume that
the dilaton, $\phi$,  entering the
action (3) is constant. We recall that a  four
dimensional action
admits solitonic string solution [8,9] when the backgrounds are such
that $\phi=$ constant, $H_{\mu \nu \rho}=0$, ${\cal F}^{i}_{\mu
\nu}=0$ and the metric as well as the moduli depend only on two
coordinates. Such a theory is an
effective two dimensional theory. Recently, Bakas [10] has considered
a four dimensional effective action with $\delta c=0$, where
$\delta c$ is the central charge deficit. One
can interprete that the action arises from compactification of a
string effective action in critical dimensions through
dimensional reduction where $M$ and ${\cal F}^{i}_{\mu \nu}$ are
set to zero ( see eq.(3)). Furthermore, the axion ( arising from
duality transformation on $H_{\mu \nu \rho}$) and the dilaton
can combined to define a complex field which transforms
nontrivially under one $SL(2,R)$. Then the existence of two
commuting Killing symmetries ( that all backgrounds depend ony
on two coordinates), is exploited to derive a form of the metric
such that the action is invariant under another $SL(2,R)$ and
the resulting theory is described by a two dimensional action. Thus,
this dimensionally reduced theory has a
symmetry which can be infinitesimally be identified with the
$O(2,2)$ current algebra [10]. In contrast, in the present
investigation, $M$,
expressed in terms of moduli $G$ and
$B$, is spacetime dependent and other backgrounds fulfill the
restrictions of constant $\phi$ and vanishing ${\cal F}_{\mu
\nu}^{i}$. The relevant action is

\bea S_{2} = \int d^2x \sqrt {-g}
\bigg\{ R +  {1 \over 8} {\rm tr} (\partial_\mu M^{-1} \partial^\mu
M) \bigg \}. \eea

\noindent Notice that, for constant $\phi$,
$\partial_{\mu}{\phi} \partial^{\mu}{\phi}$ term is absent.
Since we are considering two
dimensional spacetime, we can choose the spacetime metric
$g_{\mu \nu} = e^{\alpha (x,t)} \eta_{\mu \nu} $. Here
$\eta_{\mu \nu}$ is the flat diagonal spacetime metric $=$ diag
$(-1, 1)$ ( not to be confused with the $O(d,d)$ metric ). The
Einstein term of the action in two dimensions is a
topological term and it does not contribute to the equations of
motion. Thus the equations of motion associated with the matrix
M is of primary importance to us. It is more convenient to go
over to an $O(8,8)$ metric, $\sigma$, which is diagonal and is
related to $\eta$ by the following transformation:
$ \sigma = \rho^T \eta \rho$, where
\bea \rho = {1 \over {\sqrt 2}} \pmatrix { 1 & -1 \cr 1 & 1\cr},
\qquad  \sigma = \pmatrix {1 & 0\cr 0 & -1\cr }, \eea
\noindent and matrix
elements $1$ stand for $d \times d$ unit matrix. Then, $M
\rightarrow {\cal U } = \rho^T M \rho$ and the ${\cal U}$ satisfies
the property: ${\cal U}^T = {\cal U}$ and $\sigma {\cal U}
\sigma = {\cal U}^{-1}$. The action eq.(8) takes the form

\bea S_{2} = \int d^2x
\bigg\{ R + {1 \over 8} {\rm tr} (\partial_\mu {\cal U}^{-1}
\partial^\mu {\cal U}) \bigg \}. \eea

\noindent The equations of motion for the $\cal U$ are
\bea \partial^{\mu} {\cal A}_{\mu} = 0, \qquad {\cal A}_{\mu} =
{\cal U}^{-1} \partial_{\mu} {\cal U}  \eea
\noindent and we observe that ${\cal A}_{\mu}$  is a pure gauge. Therefore,
$[ {\cal D}_{\mu}, {\cal D}_{\nu}] = 0 $, with ${\cal D}_{\mu} =
\partial_{\mu} + {\cal A}_{\mu}$. It is worthwhile mentioning
that ${\cal A}_{\mu}$ coincides with the vector field introduced in
ref.1 to construct the infinite set of NCC. The equations of
motion (11) and the curvaturelessness proprties of $\cal
A_{\mu}$ were utilized to construct these currents by employing
the known techniques [11] for our problem.

\par The infinitesimal transformations, on the $O(d,d)$ valued
function ${\cal U}$, associated with the infinite
set on NCC are given by

\bea \delta^{(n)}{\cal U} = - {\cal U} {\Lambda}^{(n)} \eea

\noindent The  set of $\{ {\Lambda}^{(n)} \}$ are recursively defined as
\bea {\Lambda^{(n+1)}(t,x)} = \int_{-\infty}^{x} dy {\cal D}_{0}
{\Lambda^{(n)}(t,y)} = \int_{-\infty}^{x} dy \bigg\{ \partial_{0}
{\Lambda}^{(n)}(t,y) + [{\cal A}_{0}(t,y), {\Lambda^{(n)}] }
\bigg \},\eea
\noindent with ${\Lambda}^{0} \equiv T $, $T$ being a generic form of
an infinitesimal transformation of the group ${\cal G}$ and $T$ can
be expanded as a linear combination of the set $\{T_{\alpha}
\}$. Furthermore,
\bea \Lambda^{(1)} = [X_{1}, T] = \int_{-\infty}^{x} dy [{\cal
A}_{0}(t,y), T] \eea
\bea \Lambda^{(2)} = [X_{2}, T] + {1\over 2} [X_{1}, [X_{1},T]] \eea
\noindent where $X_{1} = \int_{-\infty}^{x} dy {\cal A}(t,y)$
and $X_{2}$ satisfies the equation $\partial_{1} X_{2} =
\partial_{0} X_{1} - {1\over 2} [\partial_{1}X_{1}, X_{1}]$.
\par In what
follows, we present the essential steps to construct the
generators of the Kac-Moody algebra and demonstrate the
existence of the algebra for the theory described by eq.(9).
Here we adopt an elegant and economic technique due to Devchand
and Fairlie [12] to
derive the algebra. Let us introduce the generating function for
the ${\Lambda}$'s as
\bea S({\xi}) = \sum_{0}^{\infty} {\Lambda}^{(r)} {\xi}^{r} \eea
\noindent using the recursion relation, eq.(12), and the properties of
${\cal A }_{\mu}$, we can show
\bea (\partial_{1} - \xi \partial_{0} ) S = \xi [{\cal A}_{0}, S] \eea
\noindent and $S$ can be expressed as

\bea S(\xi) = Q(\xi) T Q^{-1}(\xi). \eea
\noindent Now $Q$ satisfies the equation
\bea Q (\partial_{1} - \xi \partial_{0}) Q^{-1} = - \xi {\cal A}_{0}
 \eea
\noindent and $Q$ is defined as limit: $Q = lim_{N\rightarrow
\infty}Q_{N}$; with

\bea Q_N = e^{X_{N} {\xi}^N}..e^{X_2 {\xi}^2}e^{X_1 \xi}.  \eea
\noindent We can check by explicit calculations that
coefficients of $\xi$ and ${\xi}^2$ in (18) and (19) give us
equns. (14) and (15).
\par Moreover, it can be shown, following ref.12  that, under an
infinitesimal transformation, $\delta {\cal U} = - {\cal U} S$,
the variation of the Lagrangian density (9) is
\bea \delta {\cal L} = {1\over 4}\partial_{\mu}\epsilon^{\mu \nu}
tr[\xi {\cal A}_{\nu} + ( \xi + {1\over {\xi}}) Q^{-1} \partial_{\nu}QT]
\eea
\par In order to derive the algebra, first we define the generators of the
transformations and then evaluate commutators of two
transformations. Now,
we label each transformation with an index. For
definiteness, we choose two transformations to be
$\delta_{\alpha} {\cal U} = - {\cal U} S_{\alpha}$ and $
\delta_{\beta}{\cal U} = -{\cal U} S_{\beta}$; ${\Lambda}^{0}$
appearing in the expansions, eq.(16), for $S_{\alpha}$ and
$S_{\beta}$ are taken to be  $T_{\alpha}$ and
$T_{\beta}$ respectively and these  gerators satisfy
$[T_{\alpha}, T_{\beta}] = f_{\alpha \beta \gamma} T_{\gamma}$.
Of course, we could have chosen any two
arbitrary generator $T_a$ and $T_b$ $\in {\cal G}$; in that
case each of these generators will be expanded in terms of
the basis $ \{ T_{\gamma} \} $ and the arguments we are going
to present below will go through in that general setting too
with some extra calculations.
However, we have made
this choice here to facilitate simplicity in  computations and bring out
the essence of the arguments. Let us define following Dolan [7]
\bea {\cal M}_{\alpha}(\xi) = \int d^{2}y {\cal U}
S_{\alpha} {\delta \over {\delta {\cal U}(y)}} \eea
\noindent Then the commutator of two transformations are
\bea [{\cal M}_{\alpha}(\xi), {\cal M}_{\beta}(\zeta)] =
\int d^{2}y {\cal U} [S_{\alpha}(\xi),
S_{\beta}(\zeta)]{\delta \over {\delta {\cal U}(y)}}  -
\int d^{2} y {\cal U} [\delta_{\alpha} S_{\beta}(\zeta) -
\delta_{\beta} S_{\alpha}(\xi)]{\delta \over {\delta {\cal
U}(y)}}. \eea
\noindent The variation, $\delta_{\alpha} S_{\beta}(\zeta)$, can
be expressed as
\bea \delta_{\alpha} S_{\beta}(\zeta) = - {\zeta \over {\zeta -
\xi}} \bigg \{
[S_{\alpha}(\xi), S_{\beta}(\zeta)] - f_{\alpha \beta
\gamma} S_{\gamma}(\zeta) \bigg \} \eea
\noindent after some computations [7,12], and a similar equation
holds for $\delta_{\beta}
S_{\alpha}(\xi)$ with appropriate argument and indices.
Using the above relations in eq.(22), we arrive at
\bea [{\cal M}_{\alpha}{(\xi)}, {\cal M}_{\beta}{(\zeta)}] =
f_{\alpha \beta \gamma} \int d^{2} y {{{\cal U}[ \xi
S_{\gamma}(\xi) - \zeta S_{\gamma}(\zeta)}] \over {\xi -
\zeta}} {{\delta \over {{\cal U}(y)}}} \eea
\noindent This elegant form of equation was derived in [12].
The Kac-Moody algebra is derived as follows: Note that
${\cal M}_{\alpha}(\xi)$ can be expanded in a power series
in $\xi$ as
\bea {\cal M}_{\alpha}{(\xi)} = \sum_{0}^{\infty} {\cal
M}^{(l)}_{\alpha} {\xi}^{l} \eea
\noindent inserting the expansion eq.(27) in the commutator
(28) and comparing the coefficients of ${\xi}^{m}
{\zeta}^{n} $ on both the sides we arrive at the desired
Kac-Moody algebra
\bea [{\cal M}^{m}_{\alpha}, {\cal M}^{n}_{\beta}] =
f_{\alpha \beta \gamma}{\cal M}^{(m+n)}_{\gamma} \eea
\par A few remarks are in order here: The NCC constructed in
ref.[1] can be expressed in
terms of ${\cal U} \in O(8,8)$ and is related to the the
M-matrix: ${\cal U} = {\rho}^T M \rho$. An
arbitrary element of $O(d,d)$ can be expressed in terms of $2d^2
- d$ independent parameters. But we know that ${\cal U}$,
alternatively $M$, is determined in terms of the moduli $G$ and
$B$ and thus has only $d^2$ parameters. In fact, it was shown
by Maharana and Schwarz [2] that the moduli appearing in the
effective action, parametrize the coset $O(d,d)
\over {O(d) \otimes O(d)}$ and thus the matrix valued function
${\cal U}$ can be expanded on a basis which belong to the coset
$O(8,8) \over {O(8) \otimes )(8)}$. Indeed, the NCC were derived
in [1] by going over to the coset reformulation [2] of the effective action
(9) and then construct the curvatureless vector field ${\cal
A}_{\mu}$. Notice that if we had not set to zero the $U(1)^{16}$
gauge field action in $\hat S$ the resulting coset will be
$O(8,24) \over {O(8) \otimes O(24)}$ all our arguments will
still be valid.
Recently, it has been recognized that the string
effective actions in lower dimensions exhibit a rich symmetry
content. The dimensionally reduced effective theory ( coming
from $10$-dimensional heterotic string action with the inclusion
of  $16$ Abelian gauge fields ) in
$4$-dimensions possesses two symmetries [3]: $O(6,22;Z)$ T-duality and
$SL(2,Z)$ S-duality [13]. For $D=3$, the theory has a bigger invariance
group, $O(8,24;Z)$, and it has been shown that $SL(2,Z)$ and
$O(7,23;Z)$ T-duality are a part of this
group [14]. Now, we see that in two spacetime dimensions there
is an infinite dimensional  symmetry algebra.
\par To summarise, we have demonstrated the existence symmetrty
transformations of associated with each of the infinite
sequence of conserved currents in the two dimensional effective
theory. The generators of the infinitesimal transformations, associated with
these currents, satisfy Kac-Moody algebra which is
very intimately related with the T-duality group.

\noindent Acknowledgements: I would like to thank John H.
Schwarz and Ashoke Sen for very valuable discussions which led
to this investigation. It is a pleasure to acknowledge
stimulating discussions with the members of the High Energy
physics Group at Bhubaneswar, especially with Alok Kumar.

\def \np {{\it Nucl. Phys. }}
\def \pl {{\it Phys. Lett. }}
\def \prl {{\it Phys. Rev. Lett. }}
\def \pr {{\it Phys. Rev. }}
\def \mpl {{\it Mod. Phys. Lett. }}

\end{document}